\documentclass[apjl]{emulateapj}
\def\ergs{ergs~s$^{-1}$}
\def\ergcms{ergs$^{-1}$~cm$^{-2}$~s$^{-1}$}

\def\xa{X41.4$+$60}
\def\xb{X42.3$+$59}

\begin{document}

\submitted{Accepted for publication in The Astrophysical Journal Letters}

\title{Discovery of mHz X-ray Oscillations in a Transient Ultraluminous X-ray Source in M82}

\author{Hua Feng\altaffilmark{1}, Fengyun Rao\altaffilmark{1}, \& Philip Kaaret\altaffilmark{2}}

\altaffiltext{1}{Department of Engineering Physics and Center for Astrophysics, Tsinghua University, Beijing 100084, China}
\altaffiltext{2}{Department of Physics and Astronomy, University of Iowa, Van Allen Hall, Iowa City, IA 52242, USA}

\shortauthors{Feng, Rao, \& Kaaret}
\shorttitle{Discovery of mHz QPOs in \xb\ in M82}

\begin{abstract}
We report the discovery of X-ray quasi-periodic oscillations (QPOs) at frequencies of 3-4 mHz from a transient ultraluminous X-ray source (ULX) \xb\ in M82. The QPOs are strong and broad and appear with weak or absent red noise, and are detected only in Chandra observations when the source is brighter than $10^{40}$~\ergs. The QPO behavior is similar to the type A-I QPOs found in XTE J1550$-$564, which is a subclass of low frequency QPOs with properties in between type A and B. Therefore, we identify the QPOs in \xb\ as of type A or B, and rule out the possibility of type C. With this identification, the mass of the black hole in \xb\ can be inferred as in the range of 12,000-43,000 solar masses by scaling the QPO frequency to that of the type A/B QPOs in stellar mass black holes. Cool disk emission is detected in one Chandra observation, and the disk inner radius suggests a similar black hole mass range. Black holes of such a high mass are able to produce an energy output in a manner similar to \xb\ by accreting from the interstellar medium directly. 
\end{abstract}


\section{Introduction}

Ultraluminous X-ray sources (ULXs) are variable X-ray sources which are not coincident with the nucleus of the galaxy and exhibit luminosities over the Eddington limit of a 20 $M_\sun$ black hole ($3 \times 10^{39}$~\ergs) assuming isotropic emission. They are too bright to be powered by accretion onto stellar mass black holes like Galactic black hole binaries (BHBs) and are thus candidates of intermediate mass black holes (IMBHs). However, if the radiation is super-Eddington or beamed along the line of sight, IMBHs are not required \citep{kin01,beg02}. Modeling the X-ray spectra of ULXs could shed light on their masses. For example, the temperature and size of the accretion disk can be used to weigh the central compact object \citep{mak00}. However, identification and quantification of the disk emission in the energy spectrum, especially when it is the dominant component, has confronted difficulties and been proved unreliable in a few cases \citep{gon06,fen07a,fen09}. In contrast, characteristic time scales of the X-ray emission, such as quasi-periodic oscillations (QPOs) and the frequency break in the power spectrum, could be used to determine the mass of compact objects via a model independent calibration \citep[cf.][]{mch06,cas08}.

Low frequency QPOs from mHz to several tens Hz have been found in Galactic BHBs. On the basis of their properties like the coherence, amplitude, phase lag, and harmonic component, low frequency QPOs can be classified into three types of A, B and C \citep{wij99,rem02}. Type A/B QPOs appear in a relatively narrow frequency range at a few Hz, while type C QPOs vary in a wide frequency range in response to the spectral parameters of the source. For example, the type C QPOs in GRS 1915+105 are detected at frequencies of as low as 1 mHz \citep{mor97}. QPOs have been detected in two ULXs, and are argued to be of type C \citep{str03,str07} because they appear to vary in a manner similar to type C QPOs in Galactic BHBs.  Scaling the compact object mass with the oscillation frequency indicates the presence of IMBHs in these two ULXs \citep{fen07b,str09}. 

M82 is a starburst galaxy at a distance of 3.63~Mpc \citep{fre94}. The X-ray source \xb\ was identified as a transient ULX in M82 from multiple X-ray observations with Chandra \citep{kaa06,fen07b,kon07}. \citet{fen07b} argued that \xb\ is more likely to be an intermediate mass black hole than a stellar mass object accreting from a massive star according to its transient nature \citep{kal04}. It lies on the sky plane at $5\arcsec$ to another ULX \xa, which at most times is the brightest source in M82, and thus Chandra is the only telescope able to spatially resolve them in X-rays. \citet{kaa06} reported significant timing noise near 1 mHz from \xb\ from one Chandra observation. Here, using new Chandra data, we report the discovery of mHz QPOs in \xb\ and discuss their possible nature. 

\section{Observations and analysis}

We checked all archival Chandra and XMM-Newton observations of M82 to date, as well as three joint Chandra/XMM observations that we performed recently (PI: H.\ Feng), to search for timing noise of \xb\ at low frequencies. Among these observations, QPO features at a few mHz are found in five observations (Chandra ObsIDs 6097/8190/10027 and XMM ObsIDs 0112290201/0560590101), two out of which, 10027 and 0560590101, were made simultaneously with an over lap of 17 ks. 

%%%%%%%%%%%%%%%%%%%%%%%%%%%%%%%%%%%%%%%%%%%%%%%%%%%%%%%%%%%%%%%%%%%%%%%%%%
\begin{deluxetable*}{lccccc}[b]
\tablewidth{\textwidth}
\tablecaption{Spectral and timing properties of \xb. 
\label{tab}}

\tablehead{Instrument & XMM & Chandra & Chandra & Chandra & XMM \\
& (a) & (b) & (c) & (d) & (e)}
\startdata
Observation ID & 0112290201 & 6097 & 8190 & 10027 & 0560590101 \\
Observation date & 2001-05-06 & 2005-02-04 & 2007-06-02 & 2008-10-04 & 2008-10-03 \\
Exposure (ks) & 27 & 53 & 53 & 19 & 30 \\
$R_{\rm T}$ (counts s$^{-1}$) & 2.59 & 0.20 & 0.31 & 0.29 & 3.27 \\
$R_{\rm S}/R_{\rm T}$ & 0.7 & 1.0 & 1.0 & 1.0 & 0.2 \\
\noalign{\smallskip}\hline\noalign{\smallskip}
\multicolumn{6}{l}{Spectral Properties}\\
\noalign{\smallskip}\hline\noalign{\smallskip}
$N_{\rm H}$ ($10^{22}$ cm$^{-2}$) && $3.31 \pm 0.19$ & $3.22 \pm 0.15$ & $3.19 \pm 0.28$ &\\
$\Gamma$ && $1.44 \pm 0.09$ & $1.31 \pm 0.07$ & $1.33 \pm 0.13$ &\\
$f_{\rm X}$ ($10^{-12}$ \ergcms) && $4.49 \pm 0.10$ & $7.25 \pm 0.12$ & $6.19 \pm 0.19$ &\\
$L_{\rm X}$ ($10^{40}$ \ergs) & $2.35 - 2.47$ & $1.13 \pm 0.04$ & $1.76 \pm 0.05$ & $1.51 \pm 0.08$ &\\
$\chi^2/{\rm dof}$ && 80.7/74 & 87.1/74 & 83.6/74 &\\
\noalign{\smallskip}\hline\noalign{\smallskip}
\multicolumn{6}{l}{QPO Properties}\\
\noalign{\smallskip}\hline\noalign{\smallskip}
Significance ($\sigma$) & 3.00 & 3.39 & 6.51 & 4.49 & 2.91 \\
Frequency (mHz) & $3.98 \pm 0.18$ & $2.89 \pm 0.35$ & $3.58 \pm 0.21$ & $2.77 \pm 0.33$ & $2.97 \pm 0.33$ \\
FWHM (mHz) & $1.0 \pm 0.6$ & $1.9 \pm 0.6$ & $2.6 \pm 0.5$ & $2.0 \pm 0.6$ & $1.5 \pm 0.4$ \\
rms/mean (\%) & $5.0 \pm 0.9 $ & $10.2 \pm 1.6$ & $14.1 \pm 1.1$ & $17.4 \pm 2.0$ & $14.4 \pm 2.6$ 
\enddata
\tablecomments{$R_{\rm T}$ is the total count rate in the source extraction region, $R_{\rm S}/R_{\rm T}$ is the fraction of source photons needed for the calculation of the QPO amplitude, $N_{\rm H}$ is the absorption column density, $\Gamma$ is the power-law photon index, $f_{\rm X}$ is the observed flux in 1-8 keV, and $L_{\rm X}$ is the luminosity corrected for absorption in 1-8 keV. Errors are quoted at a confidence of 90\% for spectral parameters and 68\% (1$\sigma$) for timing parameters.}
\end{deluxetable*}

\begin{deluxetable}{ll}
\tablewidth{\columnwidth}
\tablecaption{Spectral fitting with a disk plus power-law model subject to absorption for observation 8190.
\label{tab:8190}}

\tablehead{Parameter & Value}
\startdata
$N_{\rm H}$ & $(4.2 \pm 0.4) \times 10^{22}$ cm$^{-2}$ \\
$T_{\rm in}$ & $0.17 \pm 0.03$ keV \\
$R_{\rm in}$ & $3.5_{-1.9}^{+3.0} \times 10^4$ km \\
$\Gamma$ & $1.53 \pm 0.12$ \\
$N_{\rm PL}$ & $(2.2 \pm 0.4) \times 10^{-3}$ \\
$f_{\rm X}$ & $(7.14 \pm 0.13) \times 10^{-12}$ \\
$L_{\rm X}$ & $(2.7 \pm 0.7) \times 10^{40}$ \\
disk fraction & 27\% \\
$\chi^2/{\rm dof}$ & 65.6/72
\enddata
\tablecomments{$N_{\rm H}$ and $\Gamma$ are the same as in Table~\ref{tab}. $T_{\rm in}$ is the disk inner temperature, $R_{\rm in}$ is the disk inner radius calculated assuming a face-on disk at a distance of 3.63 Mpc, $N_{\rm PL}$ is the power-law normalization in photons keV$^{-1}$ cm$^{-2}$ s$^{-1}$ at 1 keV, $f_{\rm X}$ is the absorbed flux in \ergcms, $L_{\rm X}$ is the unabsorbed luminosity in \ergs. The flux, luminosity, and the disk fraction are quoted in 1-8 keV. All errors are quoted at 90\% confidence.}
\end{deluxetable}
%%%%%%%%%%%%%%%%%%%%%%%%%%%%%%%%%%%%%%%%%%%%%%%%%%%%%%%%%%%%%%%%%%%%%%%%%%

For the three Chandra observations, the location of \xb\ on the charge coupled device (CCD) was more than $3.5\arcmin$ off the optical axis -- where a point source spreads on multiple pixels and photon pileup for sources at the flux level of \xb\ is unimportant. Event lists were created using CIAO 4.1.2 with CALDB 4.1.3. Source energy spectra in the 1-8 keV band were extracted from the source region and grouped by a factor of 4 in 1-4 keV, 8 in 4-6 keV, and 16 in 6-8 keV based on the original channels, with background subtracted from a nearby source-free region at the same location on the sky for the three observations. A power-law model subject to interstellar absorption provides adequate fits to the Chandra spectrum. The absorption column density is almost constant at about $3.3 \times 10^{22}$~cm$^{-2}$. The source spectrum is hard, with a power-law photon index of about 1.3-1.4. The luminosity corrected for absorption varied by a factor 1.5 among the three observations. All spectral parameters with 90\% errors are listed in Table~\ref{tab}. We tried to add an additional multicolor disk component to the power-law spectrum. Pronounced improvement on the fits is only obtained for observation 8190; the disk component is favored at a significance of 4.1 $\sigma$. The best-fit model parameters are listed in Table~\ref{tab:8190} with 90\% errors quoted. Power spectra were calculated from lightcurves created using events from the same region for spectral analysis with a time step equal to the CCD frame time including the readout time, which is 0.44104 s for the three observations. Each lightcurve was divided into $M$ segments with $M =$ 25, 18, and 8 for observations 6097, 8190, and 10027, respectively. A Fourier transform was performed for each segment and the individual power spectra were averaged and then geometrically binned by a factor of 1.2, 1.1, and 1.3, respectively for the three observations. The binning factor is chosen for a reasonable frequency resolution around the QPOs and a sufficiently large binning factor needed for the fitting. We checked different energy bands and found that the most significant timing noise appears in the 1-8 keV range.

For the two XMM observations, the lightcurves were produced using merged PN and MOS events in their common good time intervals (GTIs) from a half circular region \citep[region B in Fig.\ 2 of][where the contamination from the other ULX \xa\ is minimized]{fen07b} with a time step 10 times the PN frame time of about 0.73 s. Fourier transforms were performed in each GTI with 4096 points, and the power spectrum was obtained by averaging all individual ones. The number of segments, $M$, is 8 and 9 and the geometric binning factor is 1.1 and 1.2, respectively for 0112290201 and 0560590101. For observation 0560590101, MOS1 data were not included because they contain too many timing gaps. Energy spectra of \xb\ are not available from XMM data due to considerable photon confusion with nearby sources.  Spectral information for XMM 0560590101 can be obtained from the simultaneous Chandra observation 10027. For XMM 0112290201, which does not have a joint Chandra observation, the count rate of \xb\ has been resolved from surface brightness fitting \citep{fen07b}. There are four reliable spectral measurements of \xb\ with Chandra \citep[the three ones reported here and observation 5644 reported in][]{fen07b}. Taking into account the spectral variations of \xb\ with an absorption column density in the range of $(3.2-3.4) \times 10^{22}$~cm$^{-2}$ and a power-law index in 1.3-1.5, the 1-8 keV luminosity of \xb\ in XMM 0112290201 can be estimated using PIMMS as in the range of $(2.35-2.47) \times 10^{40}$ \ergs. 

\begin{figure}[t]
\plotone{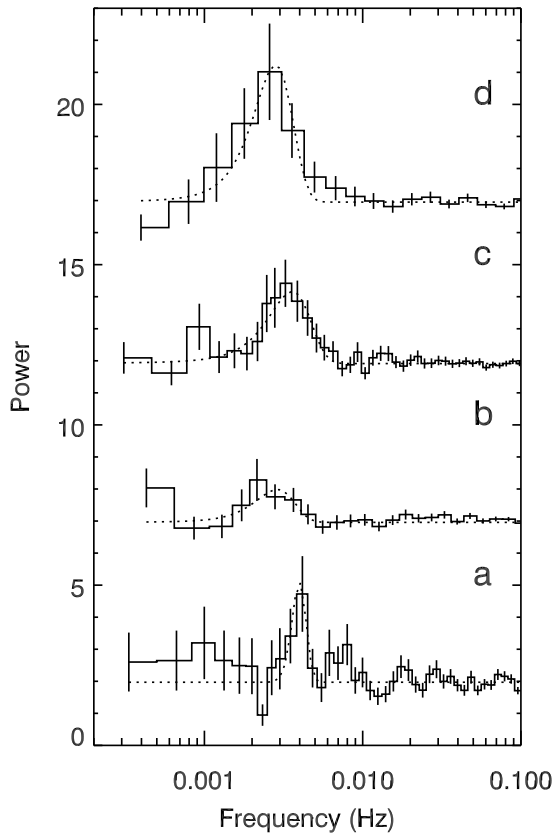}
\caption{X-ray Power spectra of \xb\ in the 1-8 keV energy range. The letters indicates the observation as shown in Table~\ref{tab}. The dotted line indicates the Gaussian plus constant model that fit the spectrum. The powers are normalized after \citet{lea83} and the baselines of b, c, and d are shifted vertically for clarity by 5, 10, and 15, respectively. 
\label{fig:pow}}
\end{figure}

Prominent timing noise above the Poisson level can be seen around a few mHz in the power spectra from these observations (Figure~\ref{fig:pow}), which we interpret as QPOs. We fit the power spectra in the whole frequency range with a a Gaussian plus constant model, which is overplotted on each spectrum with a dotted line. A Lorentzian also fits these peaks, but not as well as the Gaussian does. No obvious broadband red noise such as a (broken) power-law continuum is found near the QPO feature. We rule out the possibility that they are caused by instrumental oscillations like the dithering as follows. For the Chandra observations, photons from nearby regions do not show similar oscillations and simulations with MARX using the same aspect solution and bad pixel files confirms that no instrumentation signal is present at these frequencies. For the simultaneous Chandra and XMM observations, the same QPO feature is detected by both. The XMM power spectra are substantially contaminated by photons from the nearby variable ULX \xa, and the calculated QPO parameters may be affected. The photon fraction of \xb\ in the extraction region is estimated and taken into account when calculating the rms amplitude of QPOs from the XMM data. All QPO parameters are listed in Table~\ref{tab}, and the QPO significance is calculated based on the $\chi^2$ variation on the normalization of the QPO component. We checked the power spectra at even lower frequencies for the three Chandra observations, and found possible red noise only in 6097 appearing at frequencies well below the QPO frequency. For observation 8190, the power spectrum between 0.01-1 mHz is dominated by the white noise. For observation 10027, which is shorter than the other two, the powers between 0.1-1 mHz are also white noise dominated.

There are several other Chandra Advanced CCD Imaging Spectrometer (ACIS) observations in which \xb\ was active \citep[see][]{fen07b} but the mHz QPOs were not detected. Observation 5644 has an exposure of 70 ks and the source luminosity was $9 \times 10^{39}$~\ergs\ in 1-8 keV. The number of photons from \xb\ detected in 5644 is as many as in 6097, and twice that in 10027; the nondetection of QPOs is not due to low statistics. Observation 2933 and 6361 both have an exposure of 18 ks and the source luminosity was $4 \times 10^{39}$ and $8 \times 10^{39}$~\ergs, respectively. Another four ACIS observations (378, 379, 380-1 and 380-2) are shorter than 10 ks, and the source was slightly dimmer than in observation 2933. The mHz QPOs were not detected in any observation when the source was dim, below $\sim 10^{40}$~\ergs.

\section{Discussion}

Low frequency QPOs in stellar mass black hole X-ray binaries have a diverse population distributed in a wide frequency range. Therefore, inference of the compact object mass via the QPO frequency should be done between QPOs of the same type. The QPO coherence parameter $Q$, defined as the ratio of the central frequency to the full width at half maximum (FWHM), is around 1.4-1.5 derived from the three Chandra observations and nearly 4 from the XMM observation 0112290201. The rms amplitude of the QPOs in the three Chandra observations is high, from 10\% to 17\%, but is relatively low in the XMM observation 0112290201 of 5\%. We note that the Chandra results should be more reliable due to less contamination. As mentioned above, possible red noise is only found in Chandra 6097, and the power spectrum is totally dominated by the white noise in the other two observations at frequencies down to 0.1 or even 0.01 mHz. Therefore, the QPOs in \xb\ can be summarized as being broad and strong, and occurring with absent or weak red noise continuum. 

The QPOs in \xb\ are similar to four QPOs detected in XTE J1550$-$564 on 1999 Mar 18 and 21 and Apr 02 and 03, respectively, which also have a low $Q$ of about 1.0, an rms amplitude of about 10\%, and accompanied with a weak red noise \citep{rem02}. Another similarity is that three out of the four QPOs in XTE J1550$-$564 do not exhibit a Lorentz profile. These four QPOs were classified as `possible' type A QPOs by \citet{rem02}, or type A-I (a subclass of type A) by \citet{hom01} to distinguish them from weak type A QPOs. \citet{cas05} argued that type A-I QPOs should be classified as type B based on their frequency and amplitude. However, type B QPOs are usually narrow, with a $Q$ value of $\ga 6$ concluded from a few BHBs \citep{cas05}. Therefore, the type A-I could be a special subclass amid type A and type B. In spite of the controversy over the classification of type A-I, we identify the QPOs found in \xb\ as of type A-I based on their similar profile and the weak/absent red noise, or conservatively, of type A/B according to the discussion above. As noted above, the mHz QPOs were detected only when the source was luminous, above $\sim$$10^{40}$~\ergs. Type A/B QPOs show identical behavior that they appear only when the source flux is high enough \citep{cas04}. This further supports the identification of the mHz QPOs as of type A/B. By any means, the QPOs in \xb\ are not analogous to type C, which are narrow, strong, and accompanied by a strong red noise continuum; otherwise, the following discussion would be different as type C QPOs appear at a frequency range significantly wider than others.

The four type A-I QPOs in XTE J1550$-$564 have a central frequency varying between 7.1 to 10.3 Hz; all type A QPOs in XTE J1550$-$564 are found in a frequency range of 4.88-10.3 Hz, which covers the frequency range of type B QPOs (4.94-6.1 Hz) in the same source \citep{rem02}. The QPOs in \xb\ have a central frequency between 2.77-3.98 mHz. Comparing with the type A-I QPOs in XTE J1550$-$564, the upper and lower bounds of the frequency range have exactly the same ratio but scaled by a factor of about 2600. Based on the assumption of type A-I QPOs, the compact object mass of \xb\ can be inferred as 25,000-30,00 $M_\sun$ after adopting a mass of 9.68-11.58 $M_\sun$ for XTE J1550$-$564 \citep{oro02}. A more conservative estimate taking into account all type A/B QPOs leads to a black hole mass of 12,000-43,000 $M_\sun$ in \xb, inferred from the possibly smallest and largest ratios between QPO frequencies from the two sources. 

Type A/B QPOs are detected in several other BHBs besides XTE J1550$-$564, including GX 339$-$4, XTE J1859+226, and GS 1124$-$684 \citep[see][and references therein]{cas05}. In these sources, the type A/B QPOs show up in a similar frequency range, e.g.\ 4.5-7.49 Hz in GX 339$-$4 \citep{bel05} and 4.45-8.42 Hz in XTE J1859+226 \citep{cas04}. Type B QPOs are also detected in GRS 1915+105 varying in the frequency range of 2.44-6.84 Hz \citep{sol08}. They are broad but shown above a strong red noise component, partially similar to the QPOs in \xb. These sources also contains a similar massive black hole as in XTE J1550$-$564, but their masses are not as well constrained as in the latter \citep{rem06}. Therefore, inferring the mass of \xb\ via type A/B QPOs from these sources will lead to consistent but looser constraints. We note that all above estimates of the mass depend on a hypothesis that the frequency of type A/B QPOs is linearly scaled with the black hole mass, which, however, has never been tested. More work regarding the mass dependence of type A/B QPOs is needed.

In GX 339$-$4, all type A/B QPOs were detected when the source was in the soft intermediate state \citep{bel05}, also known as the steep power-law state defined by \citet{rem06}. In XTE J1550$-$564, the four type A-I QPOs were found in the steep power-law state as well \citep{rem02,sob00}. However, the X-ray spectrum of \xb\ is hard, seemingly inconsistent with the steep power-law state. Disk emission is detected in observation 8190, with a moderate fraction of 27\% in 1-8 keV. This disk component will contribute 40\% to the total emission below 1 MeV if the power-law component is not cut off. In the hard state, the disk contributes less than 25\% in 2-20 keV for stellar mass black holes. For more massive black holes, though there is no defined energy range, a disk fraction of 27\%-40\% seems too high for being classified as in the hard state. A varying disk component and an energetic power-law component suggest that the source may be in a state similar to the steep power-law state in stellar mass black holes. Interestingly, the best-fit disk inner radius is consistent with the gravitational radius of a black hole of 11,000-44,000 $M_\sun$, which is in agreement with the mass estimated from the QPO frequency. 

\xb\ lies about $5\arcsec$ from the kinematic center of M82 \citep{wel84} on the sky plane. Following the discussion in \citet{kaa01} and references therein, an upper limit of the compact object mass can be placed of about $3 \times 10^4$ $M_\sun$ such that the black hole will not fall onto the nucleus of the galaxy due to dynamical friction if the black hole was born at the same time of the galaxy. This upper limit is consistent with our mass estimate via QPOs.

\xb\ is coincident with an extended radio source known as 42.21+59.2, which is believed to be an H II region \citep{mcd02,kaa06} associated with a star cluster seen from infrared \citep{kon07}. Therefore, it could be ruled out that the source is a background AGN. If the disk component is true, its high bolometric luminosity of $1.2 \times 10^{41}$~\ergs\ and low temperature make it a good candidate for the ionization source of the H II region.

The outburst luminosity of the source is about $10^{40}$~\ergs. In the total 20 observations of the source \citep[16 reported in][plus Chandra 8190, 10027, 10025, and 10026]{fen07b}, \xb\ was found active in 12 of them, indicative of a duty cycle $\epsilon = 0.6$. Assuming an accretion efficiency $\eta = 0.1$, the mean mass accretion rate of the source is found to be $\dot{M} = \epsilon L_{\rm X} / \eta c^2 = 7 \times 10^{19}$ g s$^{-1}$. If the source is accreting from the interstellar medium directly, the mass accretion rate in g~s$^{-1}$ can be written as 
\begin{displaymath}
 \dot{M} = 1.4 \times 10^{11} \left( \frac{M}{M_\sun} \right)^2 
 \left( \frac{\rho}{10^{-24} \, {\rm g \, cm}^{-3}} \right)  
 \left( \frac{c_{\rm s}}{10 \, {\rm km \, s}^{-1}} \right)^{-3}_\textrm{\normalsize ,}
\end{displaymath}
where $M$ is the black hole mass, $\rho$ is density and $c_{\rm s}$ is the speed of sound of the medium \citep{fra02}. Adopting canonical values of the density and sound speed shown in the equation, a black hole mass of $2 \times 10^4$ $M_\sun$ is required to power the X-ray luminosity without a companion star. This means that if the mass estimated from the QPOs are reliable, the source does not have to be in a close binary system, but is massive enough to accrete from the interstellar medium directly. \citet{pel05} suggests that the Bondi rate may overestimate the true accretion rate for AGNs by 1-2 orders of magnitude. In this case, a corresponding higher black hole mass would be required. Conversely, if the ULX indeed lies within a star cluster, the gas density could be substantially higher, which would decrease the required black hole mass. Anyway, one should be cautious of the high uncertainty on such an estimate.

\acknowledgments We thank the anonymous referee for helpful comments and the mission planning teams of Chandra and XMM-Newton for making the observations possible. HF acknowledges funding support from the National Natural Science Foundation of China under grant No.\ 10903004 and 10978001, the 973 Program of China under grant 2009CB824800, and the Foundation for the Author of National Excellent Doctoral Dissertation of China under grant 200935. PK acknowledges partial support from Chandra grant GO9-0034X and NASA grant NNX08AJ26G.

%%%%%%%%%%%%%%%%%%%%%%%%%%%%%%%%%%%%%%%%%%%%%%%%%%%%%%%%%%%%%%%%%%%%%%%%%%

\end{document}